\documentclass[preprint,number,sort&compress]{elsarticle}

\usepackage{graphicx}

\usepackage{amssymb}

\usepackage{textcomp}

\usepackage{amsmath}
\usepackage{color}
\journal{NIM A}

\begin{document}

\begin{frontmatter}



\title{Measurement of cosmic-ray air showers with the Tunka Radio Extension (Tunka-Rex)}

\author[ISU]{P.A.~Bezyazeekov}
\author[ISU]{N.M.~Budnev}
\author[ISU]{O.A.~Gress}
\author[KIT]{A.~Haungs}
\author[KIT]{R.~Hiller\corref{cor1}}
\ead{roman.hiller@kit.edu}

\author[KIT]{T.~Huege}
\author[ISU]{Y.~Kazarina}
\author[IPE]{M.~Kleifges}
\author[ISU]{E.N.~Konstantinov}
\author[MSU]{E.E.~Korosteleva}
\author[KIT]{D.~Kostunin}
\author[IPE]{O.~Kr\"omer}
\author[MSU]{L.A.~Kuzmichev}
\author[ISU]{E.~Levinson}
\author[MSU]{N.~Lubsandorzhiev}
\author[ISU]{R.R.~Mirgazov}
\author[ISU]{R.~Monkhoev}
\author[ISU]{A.~Pakhorukov}
\author[ISU]{L.~Pankov}
\author[MSU]{V.V.~Prosin}
\author[INR]{G.I.~Rubtsov}
\author[IPE]{C.~R\"uhle}

\author[KIT]{F.G.~Schr\"oder}
\author[DESY]{R.~Wischnewski}
\author[ISU]{A.~Zagorodnikov}

\address[ISU]{Institute of Applied Physics ISU, Irkutsk, Russia}
\address[KIT]{Institut f\"ur Kernphysik, Karlsruhe Institute of Technology (KIT), Germany}
\address[IPE]{Institut f\"ur Prozessdatenverarbeitung und Elektronik, Karlsruhe Institute of Technology (KIT), Germany}
\address[MSU]{Skobeltsyn Institute of Nuclear Physics MSU, Moscow, Russia}
\address[INR]{Institute for Nuclear Research of the Russian Academy of Sciences, Moscow, Russia}
\address[DESY]{DESY, Zeuthen, Germany}

\cortext[cor1]{Corresponding author}


\begin{abstract}
Tunka-Rex is a radio detector for cosmic-ray air showers in Siberia, 
triggered by Tunka-133, a co-located air-Cherenkov detector.
The main goal of Tunka-Rex is the cross-calibration of the two detectors by measuring
the air-Cherenkov light and the radio signal emitted by the same air showers.
This way we can explore the precision of the radio-detection technique, especially for the
reconstruction of the primary energy and the depth of the shower maximum. The latter is sensitive to the mass of the primary 
cosmic-ray particles. 
In this paper we describe the detector setup and explain how electronics and antennas have been calibrated. 
The analysis of data of the first season proves the detection of cosmic-ray
air showers and therefore, the functionality of the detector. 
We confirm the expected dependence of the detection threshold on the geomagnetic angle and the 
correlation between the energy of the primary cosmic-ray particle and the radio amplitude.
Furthermore, we compare reconstructed amplitudes of radio pulses with predictions from CoREAS simulations, 
finding agreement within the uncertainties.
\end{abstract}


\end{frontmatter}


\section{Introduction}
\label{sec:intro}
Despite much progress in cosmic ray physics during the last century,
many questions, especially regarding the sources and mass composition of high-energy cosmic rays, remain unanswered. 
Established detection techniques have principal restrictions and current detectors already span 
over large areas and approach economical limits. To overcome these problems,
new detection principles are explored. One promising candidate is the radio technique. 
Already in the 1960s~\cite{Jelley1965,Allan1971} it was shown that the radio emission of 
air showers can be detected with antennas operating in the MHz frequency range. But due to technological restrictions 
at that time the interest in the technique faded quickly afterwards. Then, it experienced a renaissance 
during the last decade due to the fast advance and cheap availability of 
digital electronics and methods of signal processing~\cite{renaissance,LOPES,codalema,yakutsk}.

With an advancing theoretical understanding of the radio emission, 
time-varying transverse currents, caused by geomagnetic deflection of charged particles in the atmosphere, were established as the dominant emission
mechanism in air showers~\cite{ geomagnetic}. In addition, there is a contribution to the signal from the varying net charge, 
called Askaryan effect~\cite{askaryan,AERAAskaryan,CodalemaAskaryan,LOFARasymmetry}.

To evaluate the possible performance of a radio detector as a stand-alone device for measuring air showers or as
part of a hybrid detector in future projects, its properties and especially the achievable precision have to be
investigated in detail.

The Tunka Radio Extension (Tunka-Rex) is a radio detector for cosmic-ray air showers, which started data taking in October~2012.
It is situated in Siberia, close to Lake Baikal, at the coordinates 51\textdegree 48'35'' N, 
103\textdegree 4'2'' E at an altitude of 670\,m, on the same site as Tunka-133~\cite{T133},
an air-Cherenkov detector for air showers above 10$^{16}\,$eV.
Furthermore, the TAIGA experiment for gamma astronomy~\cite{taiga} is currently built at the site.
It consists of multiple detectors, from which Tunka-Grande~\cite{taiga} has the highest relevance for Tunka-Rex.
Tunka-Grande is a particle detector array based on former KASCADE-Grande~\cite{grande} scintillators.
Since Tunka-Rex plugs into the digitizers and data acquisition of Tunka-133 and Tunka-Grande, 
it is a relatively economic device compared to other digital antenna arrays.

In its first two seasons of operation, until 2014, Tunka-Rex was solely triggered by Tunka-133. 
Thus, all events of the air-Cherenkov detector have the radio signal recorded as well.
While this strips Tunka-Rex of one of the main advantages of a radio detector, 
i.e., its full duty cycle (the air-Cherenkov detector operates only during moonless nights with good weather), 
it provides an independent measurement of the shower parameters.
From end of 2014 on, a part of Tunka-Rex (see. Fig.~\ref{fig:map}, circle markers) is triggered by Tunka-Grande.

The main goal of Tunka-Rex is to cross-calibrate air-Cherenkov and radio measurements
to determine the achievable precision of the radio detector for 
the reconstruction of shower parameters, i.e., shower axis, energy and atmospheric depth of the shower maximum, 
which is a statistical measure of the elemental primary composition.
Tunka-133 provides these measurements with a precision of 15\% for the energy and 28\,g/cm$^{2}$ for the shower maximum~\cite{T133}.

In this paper we describe Tunka-Rex, its calibration and show first results 
on its performance. 
\section{The detector setup}
\label{sec:detector}
\subsection{Layout}
\label{sec:layout}
\begin{figure}[tb]
\center
\includegraphics[width=0.7\textwidth]{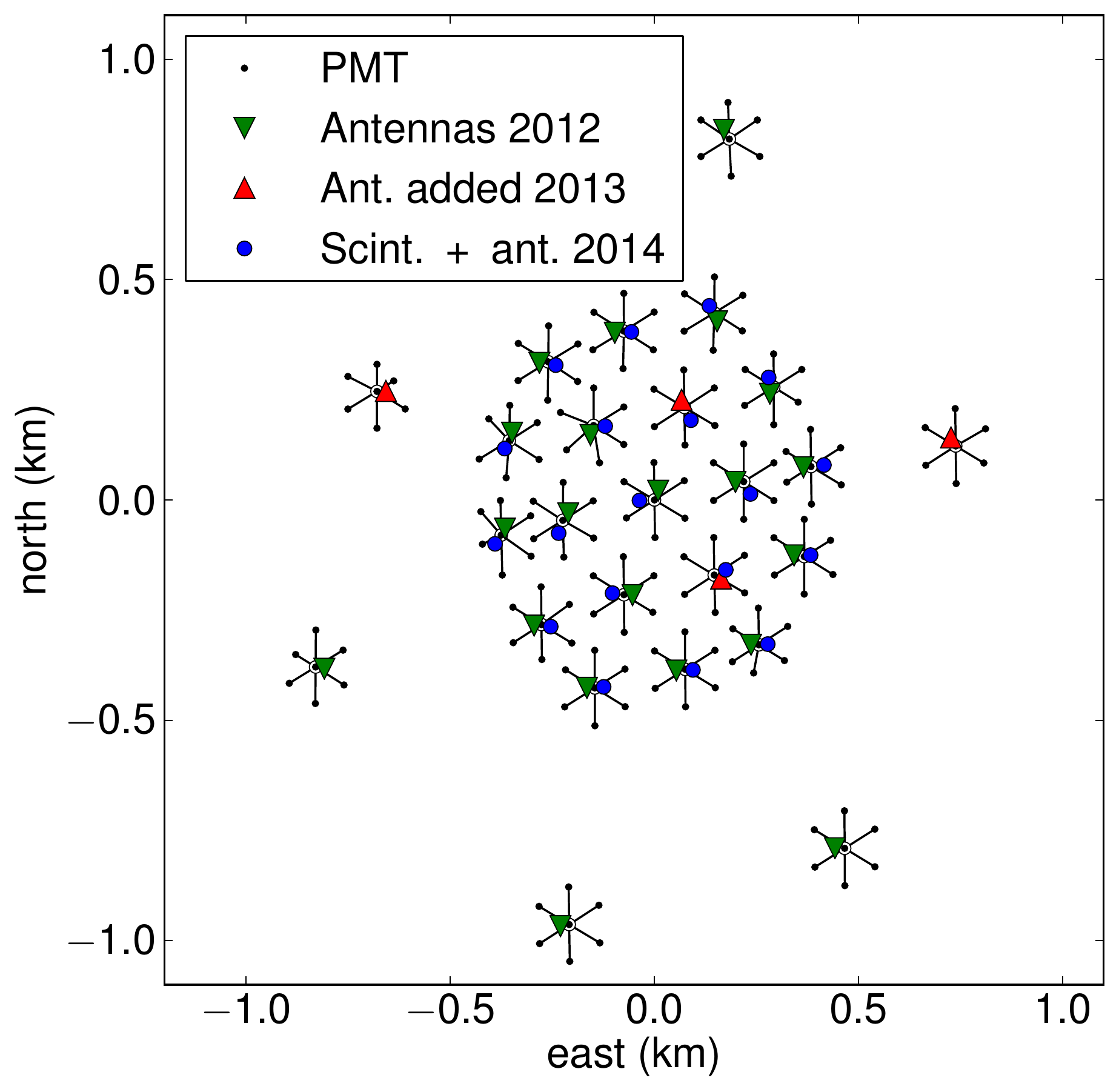}
\caption{Layout of Tunka-133 and Tunka-Rex after deployment campaigns in October 2012 and October 2013. 
In 2014, also the scintillator extension Tunka-Grande was installed with one additional antenna station at each scintillator station.}
\label{fig:map}
\end{figure} 

Tunka-133 is organized in 25 clusters, each consisting of 7 open, large-size PMTs, arranged in a hexagonal pattern and 
connected to an ADC board in its center~\cite{T133HW}. It operates only during moonless 
nights with good weather from October to April, resulting in a duty cycle of about 5\%. 
During summer it is shut down for maintenance.

The layout of Tunka-Rex is depicted in Fig.~\ref{fig:map}.
It consists currently of two separately operating detectors, with a total of 44 antenna stations:
there are 25 antenna stations triggered by Tunka-133, one per cluster of Tunka-133. 
The central, dense array of this part covers 1\,km$^2$ with 19 antenna stations, 
resulting in a spacing of about 200\,m. 
With the outer clusters a total area of about 3\,km$^2$ is reached, but the 
distance between the antennas increases in the outer region to 500\,m.{}

Additionally, there are 19 antenna stations of the same type, 
triggered by and connected to the detector stations of Tunka-Grande, featuring both surface and underground scintillators.
These antenna stations are located close to the 19 cluster centers of the inner, dense array.
They have the same design and very similar hardware as the first antenna stations.
This detector, including the antenna array, is capable of operation around-the-clock and may greatly extend 
the uptime of Tunka-Rex. Furthermore, the antenna array may be used to cross-calibrate Tunka-133 and Tunka-Grande.
This extension is still under commissioning and will not be evaluated further in this paper.

\subsection{Trigger}
\label{sec:trigger}
Until the commissioning of Tunka-Grande, Tunka-Rex was solely triggered by Tunka-133.
The trigger of Tunka-133 works on a single cluster basis~\cite{T133HW}. If at least 3 PMTs pass a threshold 
trigger within 0.5\,{\textmu}s, data from the whole
cluster is saved, including data of the Tunka-Rex antennas. 
Data from single cluster triggers are then 
combined to events in an offline coincidence search if they occur within a time window of 7\,{\textmu}s.
The radio signal is usually only detectable in events with multiple clusters above 
the detection threshold since the 
threshold for the air-Cherenkov detector is generally lower.
We apply further cuts during the reconstruction to exclude radio stations with accidental noise
(see Sec.~\ref{sec:performance}).

\subsection{Hardware of the antenna station}
\label{sec:hardware}
\begin{figure}[tb]
\center
\includegraphics[width=0.7\textwidth]{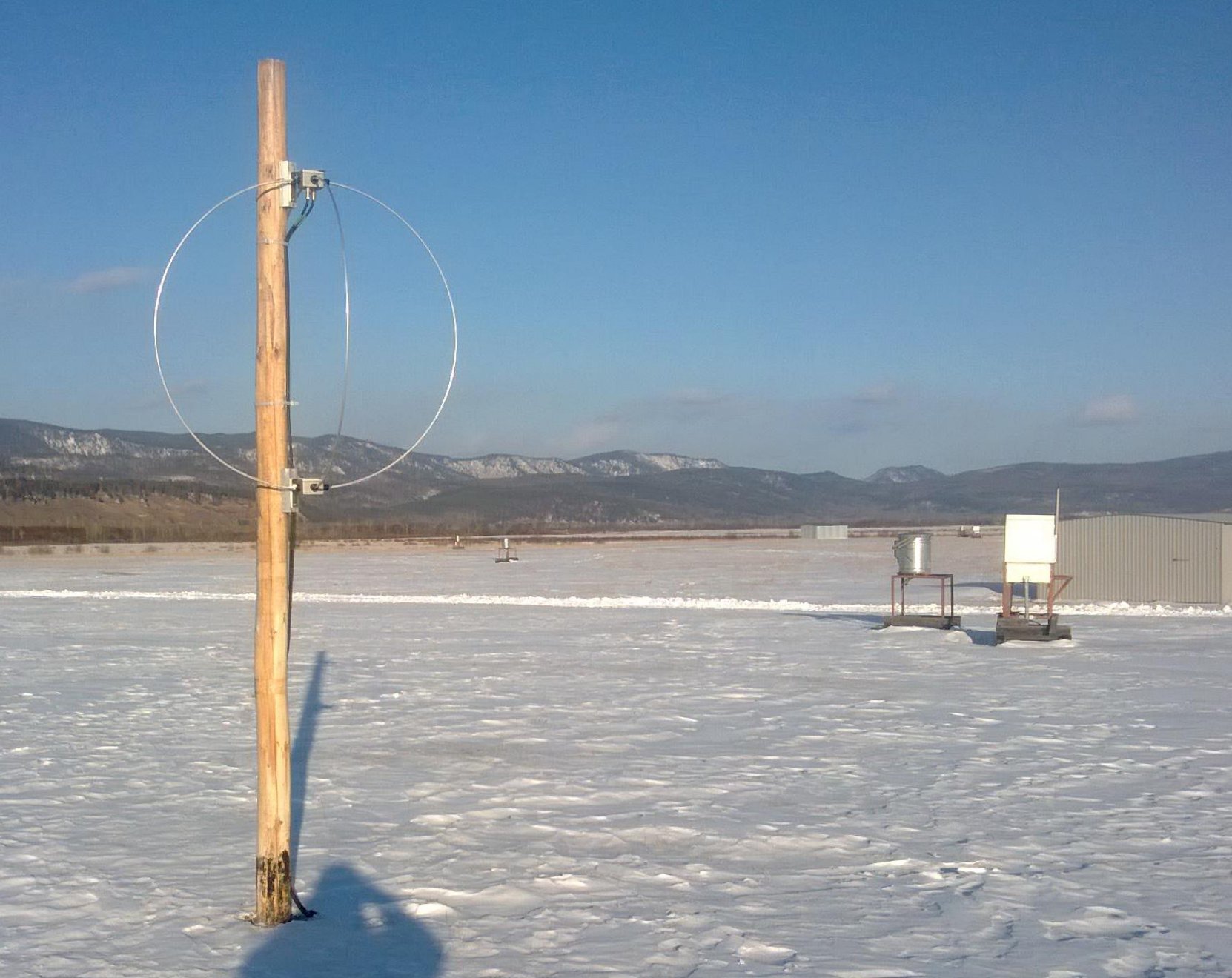}
\caption{A Tunka-Rex antenna station in front of the cluster center, housing 
filter digitization and data acquisition electronics.}
\label{fig:station}
\end{figure} 
The full hardware chain of a Tunka-Rex antenna station consists of two active antennas, 
connected via 30\,m cable to the input of a filter amplifier, which is connected to the digitization boards.

\begin{figure}[tb]
\center
\includegraphics[width=0.7\textwidth]{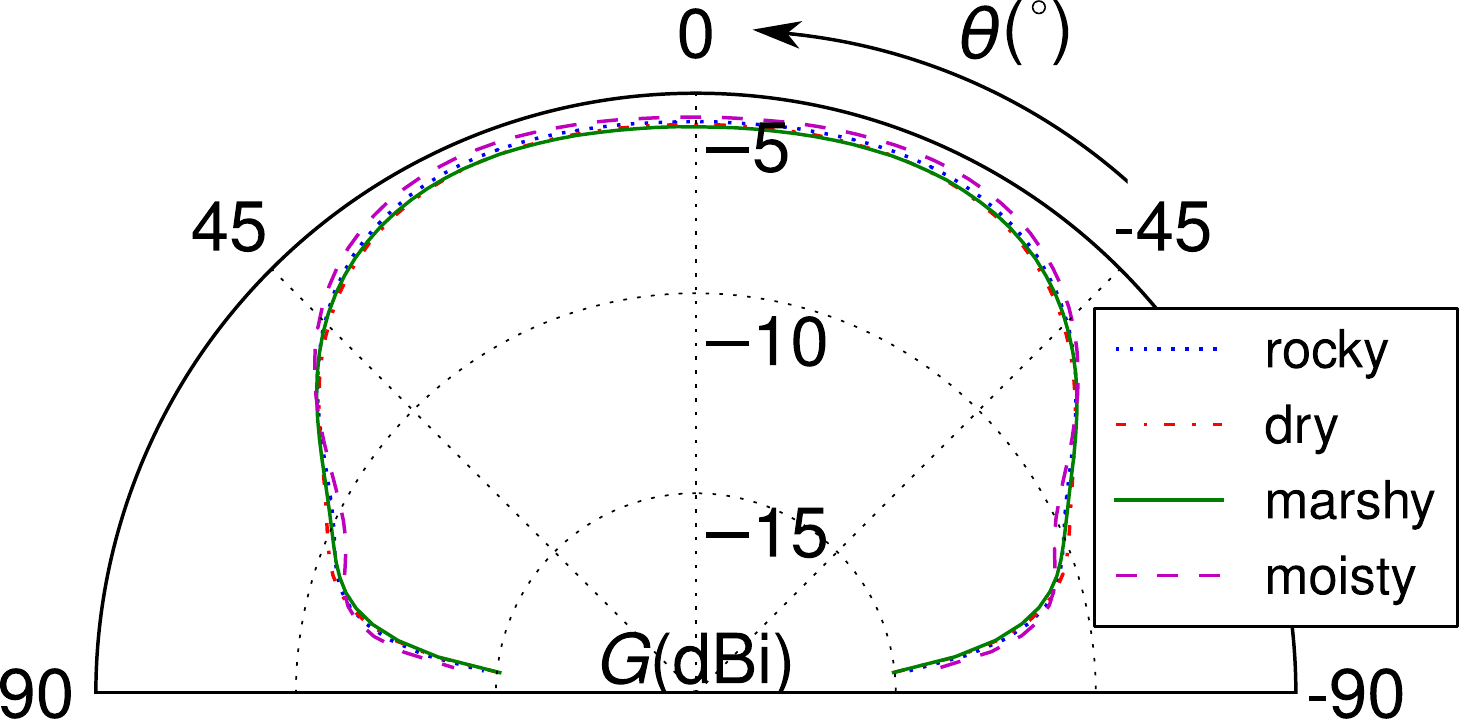}
\caption{The simulated gain G over zenith angle $\theta$ at 50\,MHz of the SALLA in the antenna-arc plane for different ground conditions. 
As expected, it has little dependence on the exact ground conditions due to the load towards ground.}
\label{fig:gainpattern}
\end{figure}

The antenna type is a SALLA (Short Aperiodic Loaded Loop Antenna)~\cite{KroemerICRC2009} 
with 120\,cm diameter (see Fig.~\ref{fig:station}), 
an economic and rugged type of loop antenna. The lower box, which 
connects the antenna arcs, houses a load, 
making the antenna less sensitive to radiation from below and thereby reducing its 
dependence on ground conditions. As shown in Fig.~\ref{fig:gainpattern} this is confirmed by an antenna simulation.
The trade-off for this lower systematic uncertainty is a generally 
lower gain compared to alternatives~\cite{AERAantennaPaper2012}.
However, the impact of the lower gain is manageable, 
since for any gain the galactic noise inevitably remains a
significant contribution to the background, limiting the possible benefit of a higher gain.
Even for the SALLA, the galactic noise is responsible for about half the background.
For a typical alternative antenna with 10\,dB higher gain than the SALLA, e.g., the energy threshold for air showers would decrease only by about 30\%.
Additionally, the loss in sensitivity is compensated by the comparably high geomagnetic field of 60\,{\textmu}T at the Tunka site, 
enhancing the dominant mechanism for radio emission in air showers.
For comparison, the geomagnetic field  is about 20\% weaker in the Netherlands, where LOFAR is located, and 60\% weaker in Argentina, at the AERA site~\cite{IGRF}.

With the half-power beam width reaching up to 140$^{\circ}$, 
the SALLA is well suited to observe a large fraction of the sky and 
covers the the full angular range of the Tunka-133 standard analysis, 
which extends up to a zenith angle of 50$^{\circ}$.

Each station has two perpendicular antennas, enabling a reconstruction of the full electric
field vector, assuming that it is perpendicular to the shower axis.
The antennas are aligned 45$^{\circ}$ rotated 
to the magnetic East-West axis.
The initial idea behind this alignment was to get a higher rate of events 
with significant signal in both channels, 
due to the predominantly East-West polarized signal~\cite{geomagnetic}, 
for the price of a few events with signal in only one channel.
However, an analysis of the distribution of geomagnetic angles revealed this effect 
to be only significant for events which are more vertical than the Earth's magnetic field.
Since the geomagnetic inclination in the Tunka valley is 72$^{\circ}$ (corresponding to a zenith angle $\theta$ of 18$^{\circ}$) 
and we are mainly sensitive to events with zenith angles $\theta>30^{\circ}$ (see Sec.~\ref{sec:performance}), 
the antenna alignment has negligible impact on our measurements.

The antennas are connected directly to a low noise amplifier in the upper antenna box (see Fig~\ref{fig:station}).
The antenna is coupled to the amplifier via a 4:1 transformer for better impedance matching. 
The amplifier enhances the signal by approximately 22\,dB and
is based on the commercial amplifier IC MGA-62563.

After the low noise amplifier the signal transverses 30\,m of RG213 cable to an active filter 
located in the cluster center.
There, a filter-amplifier enhances signal components in the band of 30-76\,MHz by 
another 32\,dB and attenuates signal components outside of this band.
The filter is an adapted version of the filter amplifiers used by AERA at the Pierre Auger Observatory~\cite{Ruehle2014}, 
designed for a nominal band of 30-80\,MHz, due to the good trade-off between radio signal strength and background in this frequency range.

The calibration of the analog hardware chain is described in Sec.~\ref{sec:HWcalib}.
\subsection{Data acquisition and signal window}
\label{sec:DAQ}
\begin{figure}[tb]
\center
\includegraphics[width=\textwidth]{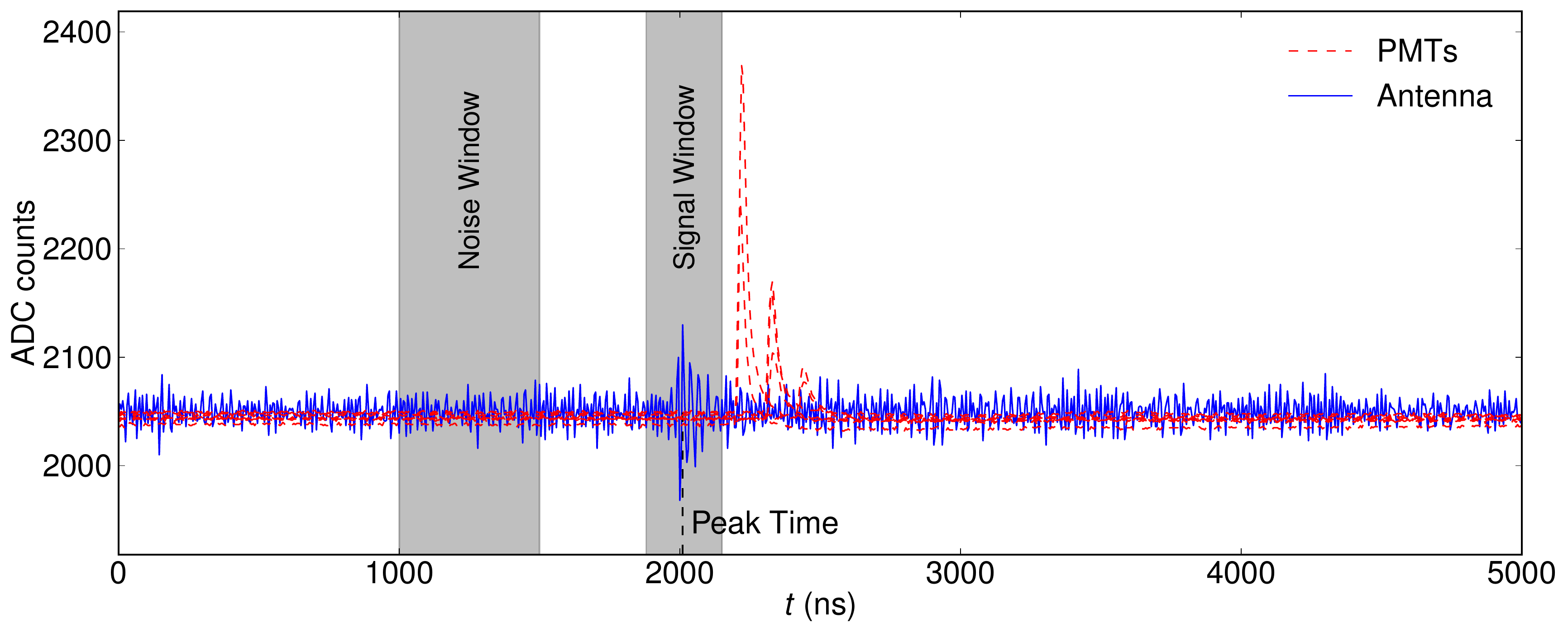}
\caption{Trace of an antenna channel with signal and the PMT traces of an air-shower event.
The antenna signal arrives before the PMT signal due to longer cabling of the PMTs.
Due to the cluster-level trigger of Tunka-133, the position of the signal window relative to the trigger time, i.e., the trace center, is always the same.}
\label{fig:trace}
\end{figure}

Finally, the signal is digitized and sampled with 12 bit depth, at 200\,MS/s, sufficient to operate in 
the first Nyquist domain for the 30-80\,MHz band. 
For each event we record a 5.12\,{\textmu}s trace (1024 samples), centered around the cluster trigger.
The exact position of the antenna-signal pulse within the trace depends on 
the relative delay in the radio-signal chain compared to the delay in the PMT-signal chain 
plus some variation due to geometry.
From the measurements mentioned in Sec.\ref{sec:HWcalib}, the delay in the radio-signal chain is 
in principle known, e.g., the electronics have a delay around 250\,ns.
Due to the longer cabling of the PMTs, we expect to observe the radio signal before the PMT signal in the trace.
Since the exact delay due to the PMTs is not known, we decided to use a phenomenological approach
to precisely determine the signal window:
we look at the time distribution of pulses in traces of antennas close to the shower axis in
high-energy events above 10$^{17}$\,eV, where we expect to see a radio signal, 
and compare them with randomly selected traces. 
In the high-energy events we observe a peak in the distribution, 
arising from radio signals of air-shower events in addition to a background floor.
The signal window is chosen to be centered around this radio signal peak with a width of 
270\,ns, containing roughly 99\% of the pulses from air showers.
In Fig.~\ref{fig:trace} an example trace of an event with indicated signal window is shown.
\subsection{Timing calibration}
\label{sec:timing}
For the offline event merging of the data from individual clusters, 
the local cluster DAQ has to be synchronized, since it operates on local clocks.
To synchronize all local cluster clocks, first, a signal is sent from the central DAQ and
the return time is measured to account for propagation delays. 
Then, the local cluster clocks receive a central reset signal at the beginning of the run.
Afterwards, a central clock signal from the central DAQ is used to form local clock signals, 
from which also the ADC sampling frequency of 200\,MHz is derived~\cite{T133HW}. 
This assures equal sampling rates and synchronization of clocks in all clusters.

To test the timing precision for the whole Tunka-Rex antenna array,
a reference beacon from LOPES~\cite{LOPESbeacon} was installed at the site and run for several nights.
The beacon emits continuous sine waves. 
Variations in the relative timing between different antennas are reflected in the recorded phasing of the beacon signals.
The analysis of the corresponding data confirmed the timing to be very stable within each night with fluctuations below 1\,ns.
However, from night to night, the relative offsets between antennas can vary up to several ns, 
leading to an overall relative timing accuracy in the order of 5\,ns \cite{PISA}.
This is sufficient for direction reconstruction of air showers not to be limited by the relative timing.
Possible improvements of the timing accuracy might still be of advantage 
for the reconstruction of the radio wavefront and application of interferometric analysis techniques~\cite{lopeswavefront}.
\section{Signal reconstruction}
\label{sec:signalreco}
For the comparison of Tunka-Rex results with other experiments or model calculations, 
the electric field strength has to be measured in absolute, detector independent units.
Therefore, the detector response has to be determined, so that the measured voltage can be converted
to the corresponding electric field strength.

The transition of the incoming signal through the analog chain can formally be described by convolution 
with the signal-chain response.
In the frequency domain the convolution 
becomes a simple point-wise multiplication.
For the two channels of one station this is in spherical coordinates:
\begin{equation}
\label{eq:antenna_trafo}
\begin{pmatrix}
 V_{1}\\
 V_{2}
\end{pmatrix} = 
\begin{pmatrix}
 H_{1 \theta}' & H_{1 \varphi}'\\
 H_{2 \theta}' & H_{2 \varphi}'
\end{pmatrix} \cdot
\begin{pmatrix}
 E_\theta \\
 E_\varphi
\end{pmatrix}
\end{equation}
With the Fourier transformed, measured voltages in the respective channels $V$, 
the antenna and electronics response $H'$, and the electric
field of the incoming signal $E$.
Due to the choice of spherical coordinates with the antenna in its center, the radial component $E_{r}$ vanishes.

From the relative timing of signal arrival in at least 3 antenna stations, the incoming direction can be reconstructed.
Eq.~\ref{eq:antenna_trafo} can then be inverted, 
if the response $H'$ is known, e.g., from calibration measurements or simulations:
\begin{equation}
\label{eq:inversion}
\vec E(\nu) = H'^{-1}(\nu) \cdot \vec V(\nu) .
\end{equation}
In real measured traces we have additional contributions from noise, which disturb the inversion.
Thus, we require a minimum signal-to-noise ratio and digitally filter contributions outside
the realized frequency band of 30-76\,MHz.

\begin{figure}[tb]
\center
\includegraphics[width=0.7\textwidth]{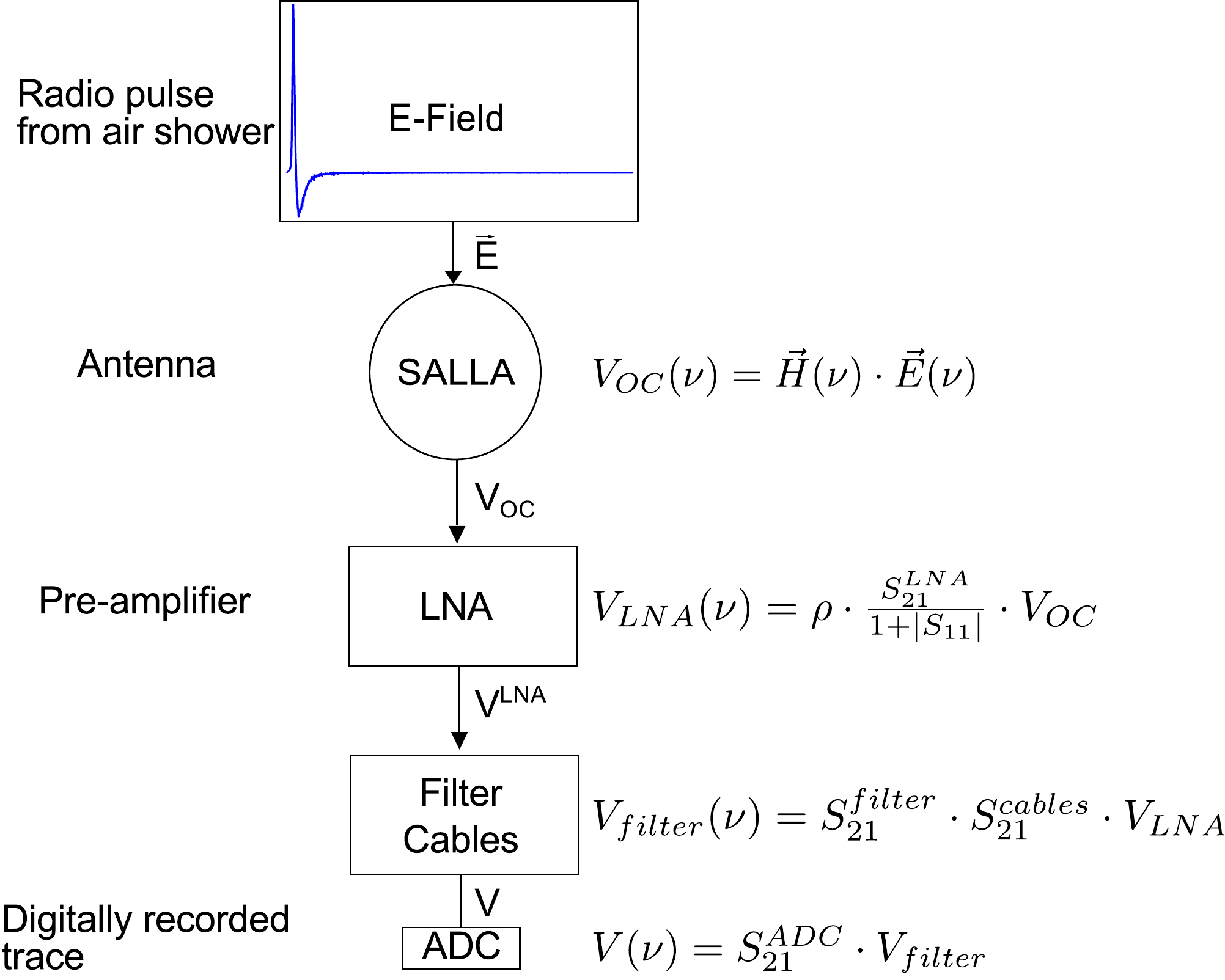}
\caption{The signal chain of Tunka-Rex, with the corresponding transformations in the frequency domain.}
\label{fig:signalchain}
\end{figure}

As illustrated in Fig.~\ref{fig:signalchain}, 
the signal-chain response can be combined from the individual
responses of the single components. 
To obtain the full response from the calibration of the full analog chain,
the ADC board and each electronics part, we decompose it into the vector effective 
length $\vec H$, 
the matching factor $\rho$ and the electronics transmission $S_{21}$:
\begin{equation}
\label{eq:electronics_decomposed}
\vec H'(\nu) = S_{21}\cdot\rho\cdot \vec H.
\end{equation}
The vector $\vec H'(\nu)$ represents one row of the the matrix $H'$ of Eq.~\ref{eq:antenna_trafo}.
\subsection{Electronics transmission parameters}
\label{sec:HWcalib}
\begin{figure}[tb]
\center
\includegraphics[width=0.7\textwidth]{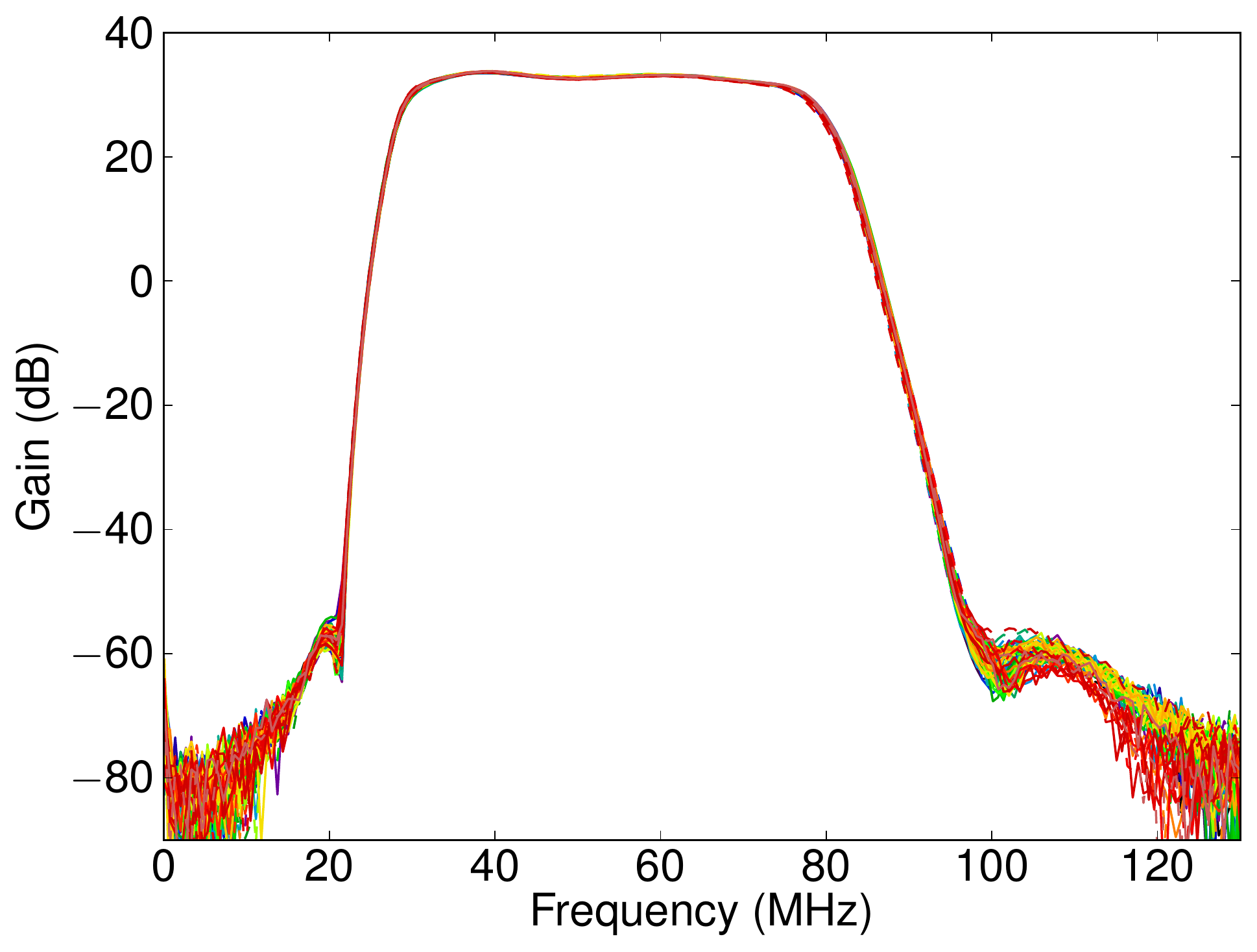}
\caption{The gain of the filters deployed at the first 25 antenna stations, measured with a network analyzer. 
Different colored lines correspond to the two channels of each individual filter.}
\label{fig:filter}
\end{figure}

The forward transmissions $S_{21}$ of the electronics parts was measured with a network analyzer
for all filters and amplifiers.
In Fig.~\ref{fig:filter}, the amplitudes of $S_{21}$ are shown for the filters of 
the first production series, deployed until 2013. 
The new production series deployed in 2014 is very similar, but has slightly different gain
and bandwidth.

\begin{figure}[tb]
\center
\includegraphics[width=0.7\textwidth]{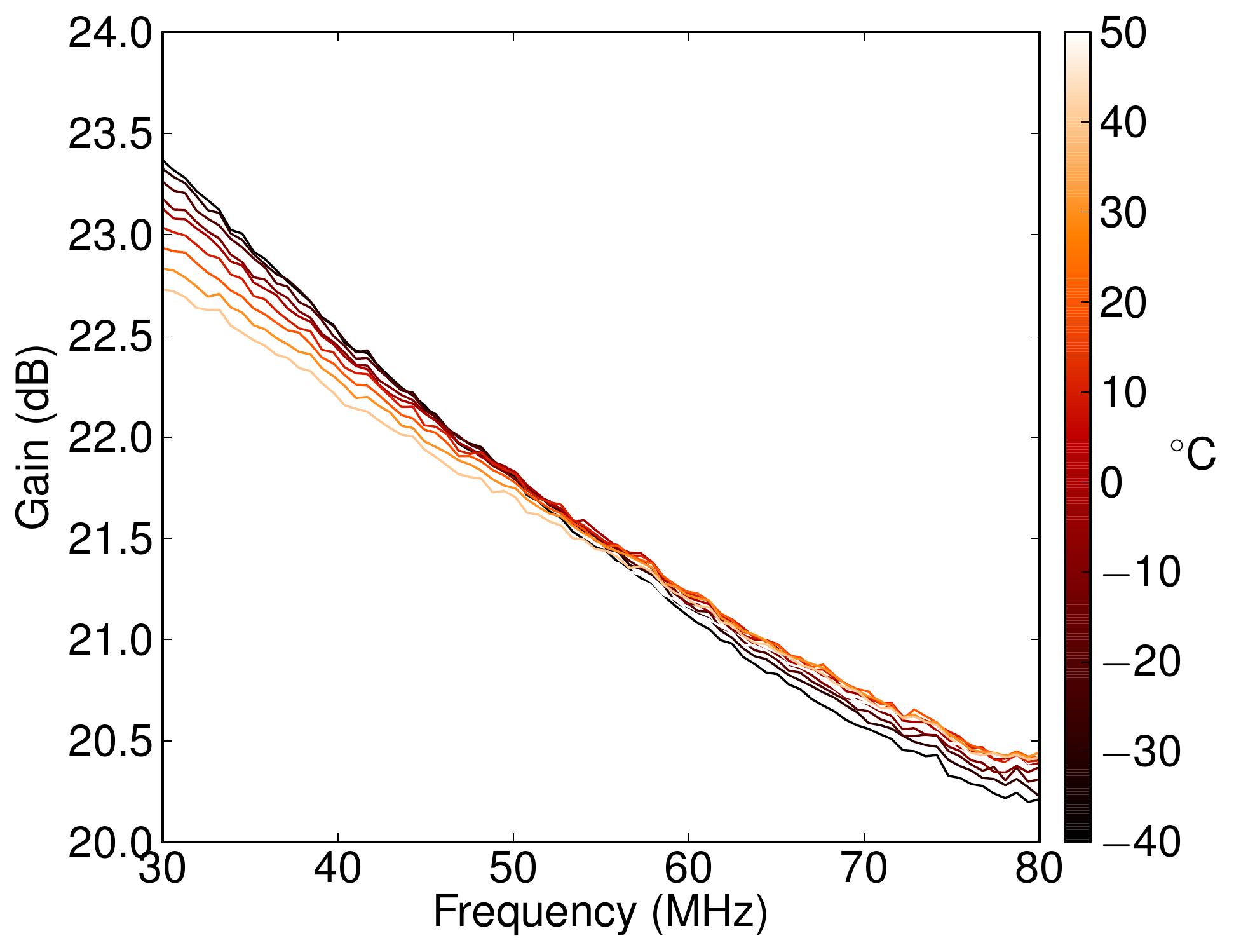}
\caption{Gain of a low noise amplifier, measured at different temperatures.
Its variation causes an uncertainty of about 6\% on the reconstructed amplitudes in the expected temperature range.}
\label{fig:temp}
\end{figure}

We also investigated the temperature dependence of the electronics in a temperature chamber, 
as the night time temperature during winter typically ranges from $-35\,^{\circ}$C to $0\,^{\circ}$C at the site.
In Fig.~\ref{fig:temp} one curve from a low noise amplifier is shown at different temperatures. 
While the filters in the cluster center are heated, the amplifiers in the antennas are
exposed to the temperature outside.
Since the temperature is not monitored for the air-shower measurements, 
it causes a systematic uncertainty in the amplitude reconstruction of about 6\%.

The transmission parameters are measured within a 50\,$\Omega$ system, 
which is a good approximation for the electronics chain, but not for the antenna.
Therefore, we introduce a correction factor $\rho$, which describes 
the fraction of the open-circuit voltage between the antenna terminals, dropping over the amplifier input.
From the Thevenin equivalent diagram with the amplifier input impedance and the SALLA 
connected in series, we determine it to be 
\begin{equation}
\label{eq:matching}
\rho(\nu) = \frac{Z_{\mathrm{LNA}}}{Z_{\mathrm{SALLA}}+Z_{\mathrm{LNA}}} .
\end{equation}
The input impedances of the amplifier and the antenna are determined from
their input reflection $S_{11}$, measured with a network analyzer:
\begin{equation}
\label{eq:impedance}
Z_{\mathrm{in}}(\nu)=Z_{\mathrm{NWA}}\cdot\frac{1+S_{11}}{1-S_{11}}\ \mathrm{with}\ Z_{\mathrm{NWA}}=50\,\Omega.
\end{equation}
Finally, we have to take reflections during the measurement of the 
forward transmission of the low noise amplifier into account:
\begin{equation}
\label{eq:S21_norm}
S_{21}'(\nu)=\frac{S_{21}}{1+\left|S_{11}\right|}.
\end{equation}
\subsection{Antenna calibration}
\label{sec:AntennaCalib}
\begin{figure}[p]
\center
\includegraphics[width=0.7\textwidth]{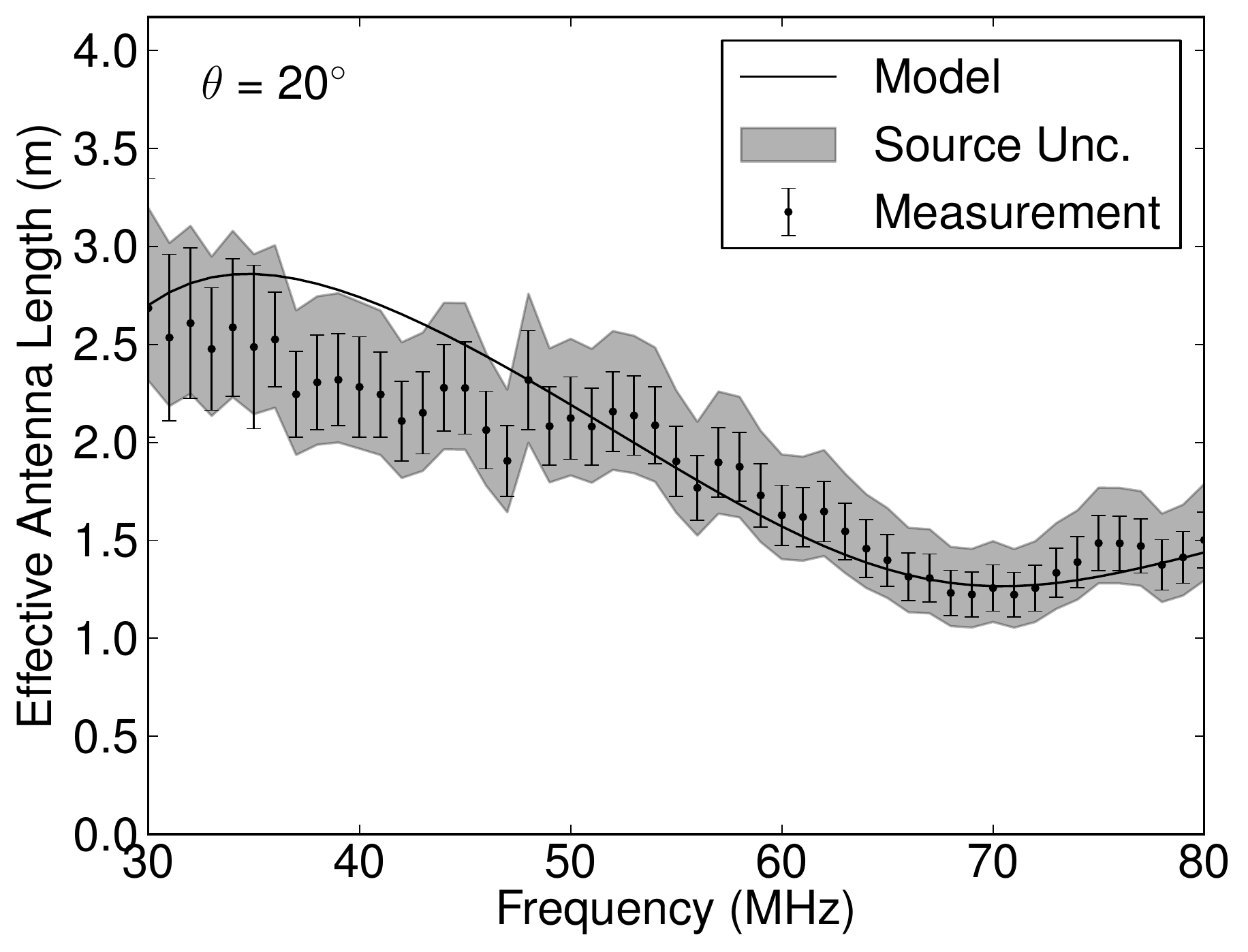}
\caption{Vector effective length for the polarization along the antenna 
for a source at a zenith angle of 20$^\circ$, obtained from NEC2 simulations and from the 
performed calibration.}
\label{fig:correction}
\end{figure}
\begin{figure}[p]
\center
\includegraphics[width=0.7\textwidth]{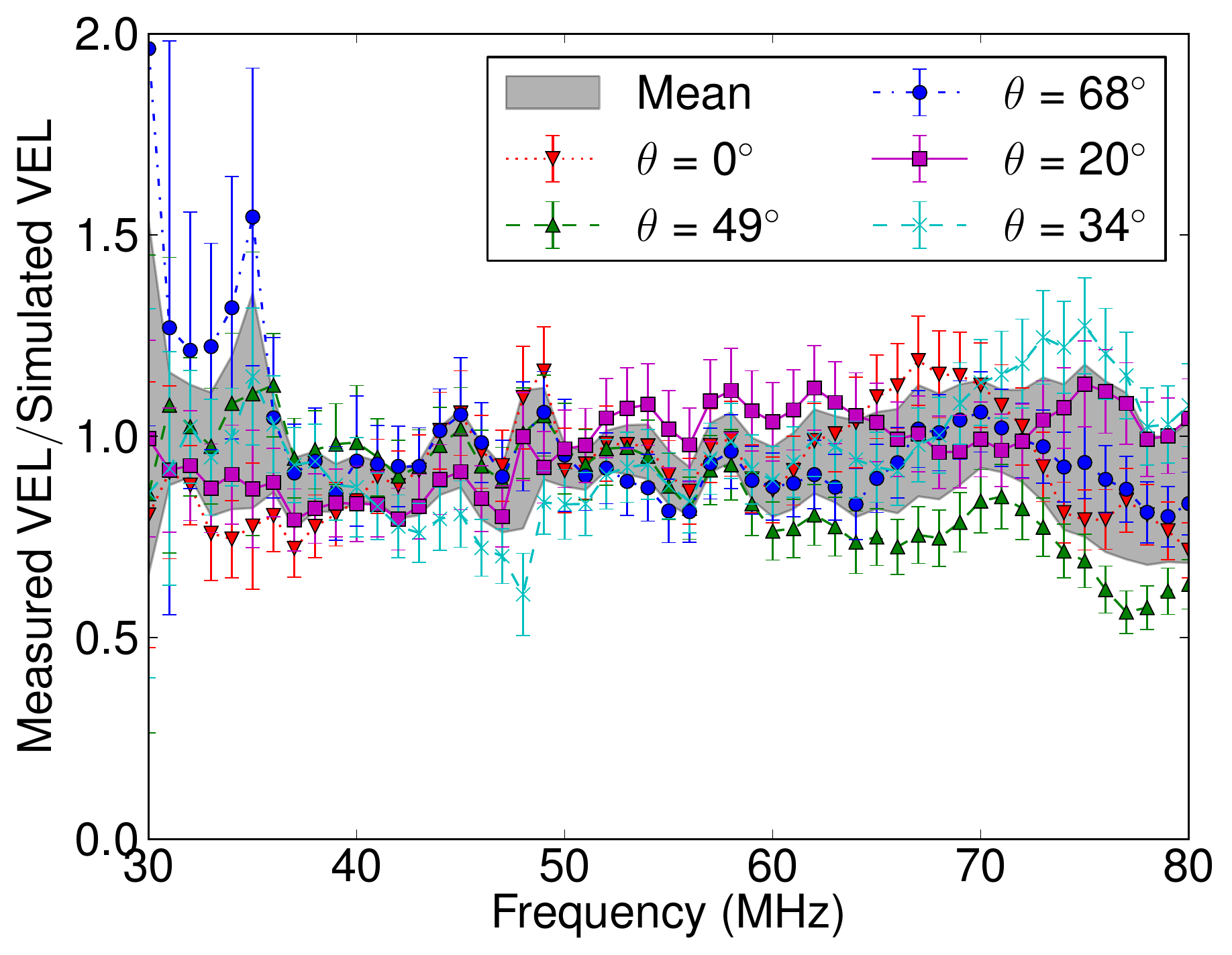}
\caption{Relative deviation between the simulated antenna model and the calibration measurements for different zenith angles.
The mean variation over different zenith angles is 12\%.}
\label{fig:factors}
\end{figure} 
To determine the antenna pattern, i.e., the vector effective length of the antenna as a function 
of incoming direction and polarization, one can either 
rely on antenna simulations or measure it with an antenna calibration technique.
Since the measurement for all solid angles and polarizations is very cumbersome, 
we combined both methods: 
we simulated the antenna pattern with the NEC2 simulation code~\cite{nec2} 
to retrieve the directivity and then performed an amplitude
calibration to normalize the simulation to.

The simulation was performed for an open ended, 1.2\,m diameter wire antenna, 
with a 390\,$\Omega$ load towards ground, resembling the SALLA.
In Fig.~\ref{fig:gainpattern}, a slice of the simulated gain pattern is shown 
for different grounds.

For the amplitude calibration a SALLA was deployed at KIT Karlsruhe, Germany. 
The only relevant difference between it and its counterparts in Russia is the ground, 
which has little impact, as shown by the antenna simulations (see Fig.~\ref{fig:gainpattern}).
The commercial reference source VSQ1000+DPA4000 by Schaffner Electrotest 
GmbH (now Teseq) was used. It is a biconical antenna, which produces a calibrated, 
linearly polarized frequency comb with 1\,MHz steps in the range 30-1000\,MHz. 
It is the same reference source, as used for the LOPES calibration~\cite{lopescalib} and 
later for the LOFAR calibration~\cite{LOFARcalibration}.
Therefore, these three experiments will have a consistent absolute calibration scale.

For the calibration of Tunka-Rex we obtained new calibration measurements for the reference source from the manufacturer.
Contrary to the old calibration of the reference source, the new one was performed for free field conditions,
better fitting our calibration situation for the reference source being situated above the antenna.
The values of the new calibration measurements were already used for an update of the LOPES calibration \cite{LOPESnewcalibration} and for the LOFAR calibration, which was successfully cross-checked against galactic background \cite{LOFARcalibration}.

The calibration procedure works as follows:
the reference source is put in about 13\,m height by a crane parking in
20\,m distance from the antenna, using a wooden carrier as a spacer.
To have measurements for incoming signals from different zenith angles, the reference antenna is 
placed directly above the antenna, as well as horizontally displaced.  
A differential GPS is used to determine 
the exact distance and angle between the calibration source and the SALLA.
The zenith angles at which the calibration was performed were $0^{\circ}$, $20^{\circ}$, $34^{\circ}$, $49^{\circ}$ and $68^{\circ}$.
The reference antenna is aligned with strings from the ground along the plane of the SALLA.
Since the original data acquisition system (DAQ) could not be brought to Karlsruhe, 
an oscilloscope was used to record traces.
Thus, the calibration is almost end-to-end, except for the final step, 
i.e., the data acquisition, which will be discussed in Sec.~\ref{sec:ADC}.

With this measurement, Fourier coefficients can be calculated from the recorded 
traces to retrieve the voltage amplitude $V$ at each frequency. 
To avoid spectral leakage, the number of points in the trace is chosen to be 
an integer multiple of $\nu_s/\Delta \nu$, with the sampling frequency $\nu_s$ and 
the frequency interval $\Delta \nu$ of the peaks (=1\,MHz). Thus, the Fourier coefficients 
lie precisely on the 1\,MHz peaks of the calibration source.

According to simulations and confirmed by the calibration measurements, 
the antenna has a sensitivity pattern close to the one of a horizontal dipole in the antenna-arc plane.
Therefore, it has maximal sensitivity for radiation which is polarized horizontally and lying in the antenna-arc plane. 
With the polarization axis of maximal sensitivity $\vec p$, the components of Eq.~(\ref{eq:antenna_trafo}) become:
\begin{equation}
\label{eq:calib}
V_{i}(\nu) = \left| \vec H'_{i}(\nu)\right| \cdot \vec p_{i} \cdot \vec E(\nu).
\end{equation}
In our case we aligned the E-field polarization with $\vec p_{i}$, so Eq.~(\ref{eq:calib}) can be solved for $\vec H'$
\begin{equation}
\label{eq:calib1}
\vec H'_{i}(\nu)= \frac{V_{i}(\nu)}{\left|\vec E(\nu)\right|}\cdot \vec p.
\end{equation}
We take $E=E_{\mathrm{calib}}\cdot r_{\mathrm{ref}}/r$ at a reference distance 
of $r_{\mathrm{ref}}=10\,\mathrm{m}$ from the calibration sheet of the reference source and 
$\vec p$ from the antenna simulation to obtain $\vec H'$ from the measurement of $V$.
$H$ can then be retrieved from $H'$ with Eq.~(\ref{eq:electronics_decomposed}).

In Fig.~\ref{fig:correction}, an example for the simulated and measured vector effective length is shown.
Overall, the simulation fits the measured profile within the uncertainties.
The variation over different zenith angles is 12\% on average over all relevant frequencies,
which we use to estimate how well the simulated directivity describes the actual pattern.

The relative deviations for all zenith angles are shown in Fig.~\ref{fig:factors}.
On average the simulated vector effective length is 6\% smaller than the measured one, 
which is consistent with the 2.5\,dB two sigma uncertainty of the source calibration 
given by the manufacturer, corresponding to 16\% (one sigma) uncertainty on the amplitude.
Furthermore, for this measurement additional systematic uncertainties have to be considered, 
e.g., several 10\% from a measurement adapter for the LNA. 
However, for the amplitude reconstruction, these systematics cancel out by including the full analog chain in the antenna calibration, thus, defining the absolute scale by the reference source calibration.
The remaining role of the calibration of individual components is to compensate production fluctuations.

For the amplitude calibration, we rescale the simulated vector effective lengths frequency wise according 
to the mean of the performed calibration measurements at different zenith angles.

The overall uncertainty on the absolute amplitude reconstruction is 22\%,
with a dominating contribution of 16\% from the calibration uncertainty of the reference source.
However, for the comparison to LOPES or LOFAR measurements, which used the same source for their calibration,
this contribution can be ignored.
The uncertainties for the amplitude reconstruction are summarized in Table~\ref{tab:calib}.
\begin{table}[htb]
\caption{Uncertainties for the absolute amplitude reconstruction, 
sorted by contributions affecting the signal of each antenna independently,
contributions changing from event to event, 
but highly correlated within an event and contributions from 
the calibration affecting the overall scale.
Totals are calculated by adding up the squared contributions.
}
\centering
\begin{tabular}{l l r}
\hline
Level & Origin & Uncertainty (\%)\\
\hline
{\bf Antenna-to-antenna} & antenna production & 1\\
& positioning and alignment & 2\\
& {\bf Total} & {\bf 2} \\
\hline
{\bf Event-to-event} & environmental temperature & 6\\ 
& ground conditions & 3\\
& antenna model & 12\\
& {\bf Total} & {\bf 14}\\
\hline
{\bf Absolute scale}& source calibration & 16\\
& source positioning + alignment & 3\\
& temperature during calibration & 6\\
& {\bf Total} & {\bf 17}\\
\hline
{\bf Total} &  & {\bf 22}
\end{tabular}
\label{tab:calib}
\end{table}

\subsection{Analog-to-digital-converter board}
\label{sec:ADC}
The final step in the signal chain is the digitization, after which the 
recorded traces are stored.
To calibrate the digitizer boards we used a signal generator to feed in sine 
waves with a known amplitude at different frequencies.
From the recorded traces the dynamic range is determined.
Although the dynamic range varies slightly with frequency, 
the transition from voltage to ADC counts has been measured to be linear in very good approximation. 
Therefore, its impact on the signal is described as a contribution on 
the forward transmission $S_{21}$ in Eq.~(\ref{eq:electronics_decomposed})
and inverted accordingly.

\section{Performance}
\label{sec:performance}

We started data taking in October 2012 with the antennas indicated in Fig.~\ref{fig:map}.
For the first season of Tunka-Rex, until April 2013, the Tunka-133 reconstruction, 
including energy and the atmospheric depth of the shower maximum, is known.
For the second season, 2013/2014, we look only at the shower geometry reconstructed by Tunka-133 for monitoring purposes, 
but leave the reconstructed energy and shower maximum blinded.
This way we test our reconstruction methods, especially for the shower maximum, 
on a different data set, than it was tuned on.
Therefore, we present here only results based on data from the 2012/2013 season, 
and the data of 2013/2014 will be unblinded later.

In the first season we accumulated about 280\,h of effective runtime.
In a simple first reconstruction we consider the Tunka-133 event sample
with reconstructed primary energies above 10$^{16.5}$\,eV. 
It contains only events with zenith angles $\theta\leq50^{\circ}$,
for which Tunka-133 is assumed to have full efficiency above 10$^{16}$\,eV 
and can provide a full reconstruction, including primary energy and shower maximum.

\begin{figure}[tb]
\center
\includegraphics[width=0.5\textwidth]{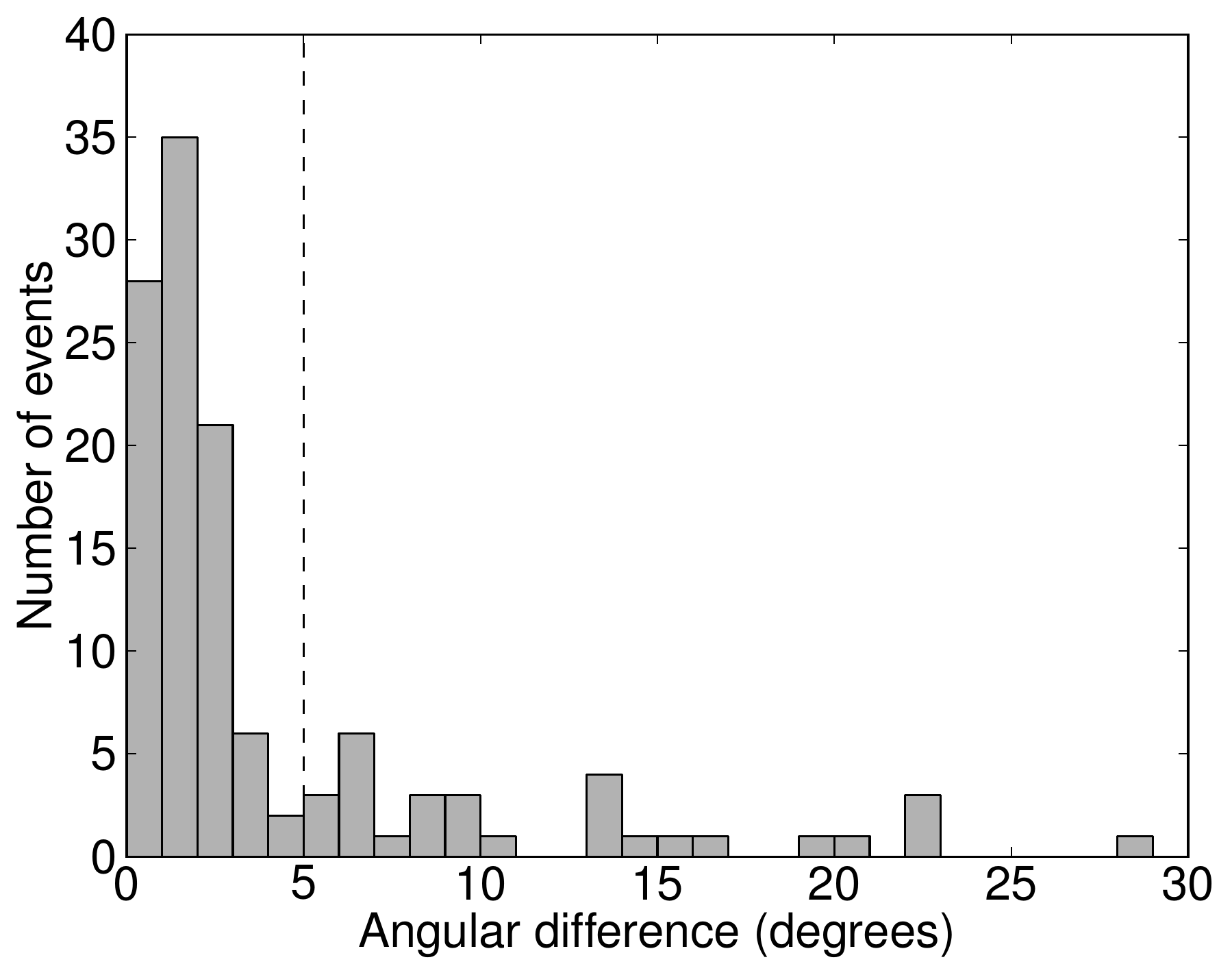}
\caption{Angular difference between the shower axis reconstructed by Tunka-133 and Tunka-Rex.
The peak around 2$^{\circ}$ can be appointed to the angular resolution of Tunka-Rex.
To select high quality events we cut events with deviations larger than 5$^{\circ}$.}
\label{fig:angle}
\end{figure}

For the analysis of the data we use a modified version of the Offline software framework~\cite{RadioOffline2011}, 
developed by the Pierre Auger Collaboration.
After reading the raw data, the hardware response is inverted. 
Then we cut the frequency range digitally 
to 35-76\,MHz, because we regularly observe
broad band noise, probably from electronics of the cluster center, 
up to 35\,MHz, and have the upper edge of the realized band of the filter 
amplifiers at 76\,MHz. 
Additionally, regularly observed narrow band interferences at full 5\,MHz steps are removed.
With the antenna model we reconstruct the electric field vector using the 
reconstructed incoming direction from Tunka-133. 
Afterwards we determine a noise level 
$N$ by taking the RMS of a 500\,ns long window shortly before the signal window, and a signal level $S$ by taking the 
peak of the Hilbert envelope in the signal window (see Fig.~\ref{fig:trace}).
All stations with a signal-to-noise ratio (SNR, $S^2/N^2$) below 10 are rejected.
If two consecutive stations in the lateral distribution (amplitude vs. distance to shower axis) do not pass the SNR cut, 
we exclude all stations which are even further away from the shower axis to remove
false positive signals due to background peaks far from the shower core.
Then we select events with at least 3 antennas passing all cuts.
After these cuts we are left with 122 events.
To exclude events with significant impact from transient background peaks, 
we determine the arrival direction from the pulse arrival times of the antennas and
require it to agree with the reconstructed direction of Tunka-133
within 5$^\circ$. In Fig.~\ref{fig:angle} the angular differences of all events are shown. 
This leaves us with 91 high quality events with full reconstruction from both, radio and air-Cherenkov detector, including energy and 
the shower maximum determined by Tunka-133.

Furthermore, to measure inclined events with zenith angles above 50$^{\circ}$, 
we release the cut on the maximum zenith angle of Tunka-133.
To account for the increased geometrical delays we increase the width of the signal window to 350\,ns
and the minimum SNR to 12, to compensate for the wider signal window.
This leads to another 85 inclined events. 
For these events Tunka-133 can only provide a reconstruction of the direction and a rough estimate of the shower core, but no reliable reconstruction 
of the energy or shower maximum.

\begin{figure}[tb]
\center
\includegraphics[width=0.5\textwidth]{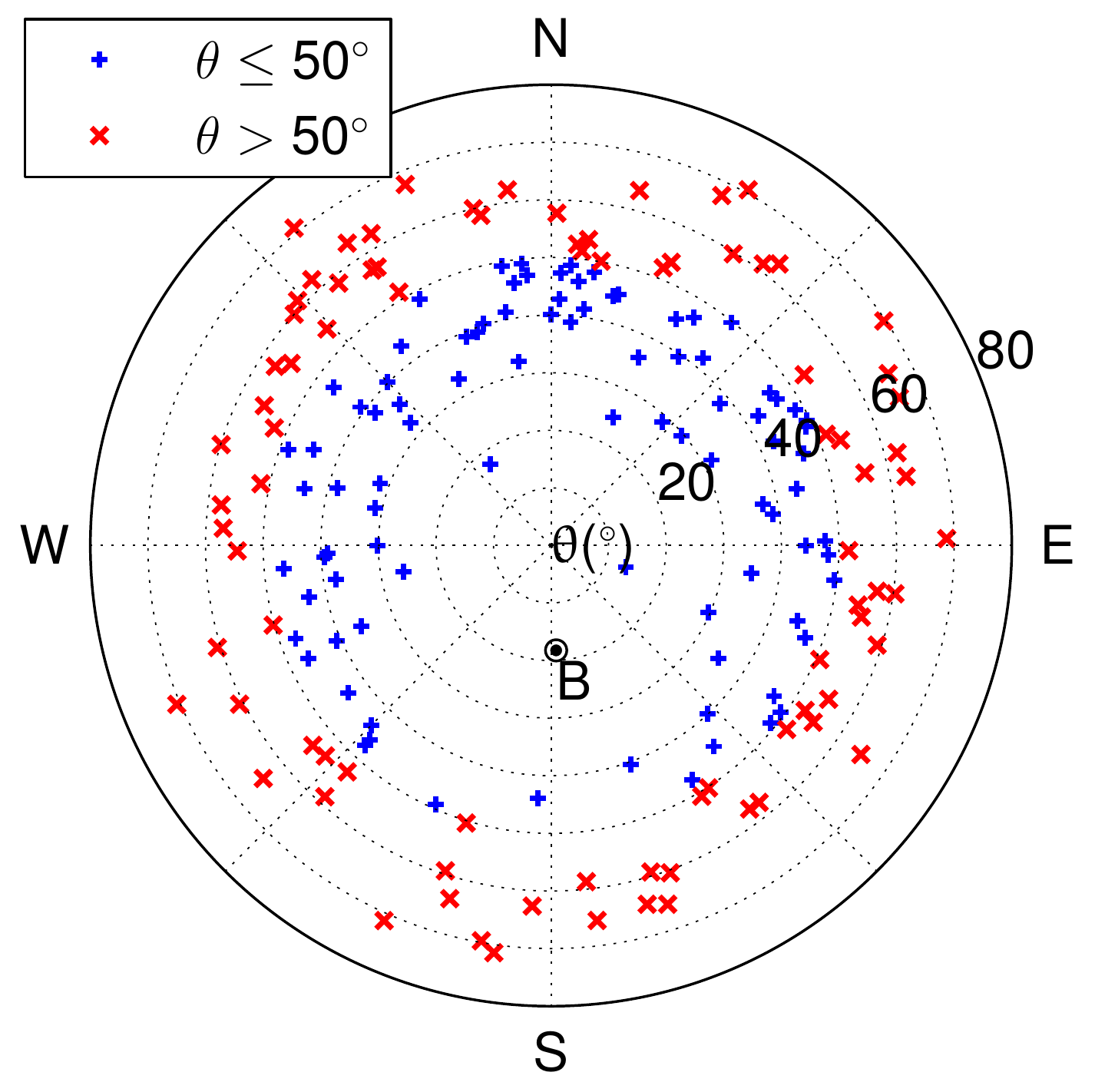}
\caption{Sky map of the event distribution of the 2012/2013 season of Tunka-Rex. 
Events close to the geomagnetic field vector have small geomagnetic angles, 
and have therefore little radio signal. Since we operate close to the 
detection threshold this leads to a north-south asymmetry 
in the event distribution.}
\label{fig:sky}
\end{figure}

In Fig.~\ref{fig:sky} the distribution of the resulting 176 events on the sky is shown. 
Looking only at the vertical events ($\theta \leq 50^{\circ}$), we observe 31 events in the Southern 
Hemisphere versus 60 events in the Northern Hemisphere. This asymmetry is
also observed by other experiments~\cite{CODALEMAasymmetry, LOPES3D, AERAasymmetry, LOFARasymmetry}. 
It originates from the dependence of the detection threshold on the asymmetric distribution of 
geomagnetic angles (angle between geomagnetic field and shower axis) on the sky.

The low abundance of events with zenith angles below 30$^\circ$ is because of the 
small footprint of the radio signal on the ground for vertical showers, combined with the antenna spacing of 
about 200\,m and the restrictive cuts, requiring 3 antenna stations above threshold.
Additionally, the geomagnetic field is nearly vertical at the Tunka-Rex site and consequently leads to
low geomagnetic angles for vertical showers.

\begin{figure}[tb]
\center
\includegraphics[width=0.7\textwidth]{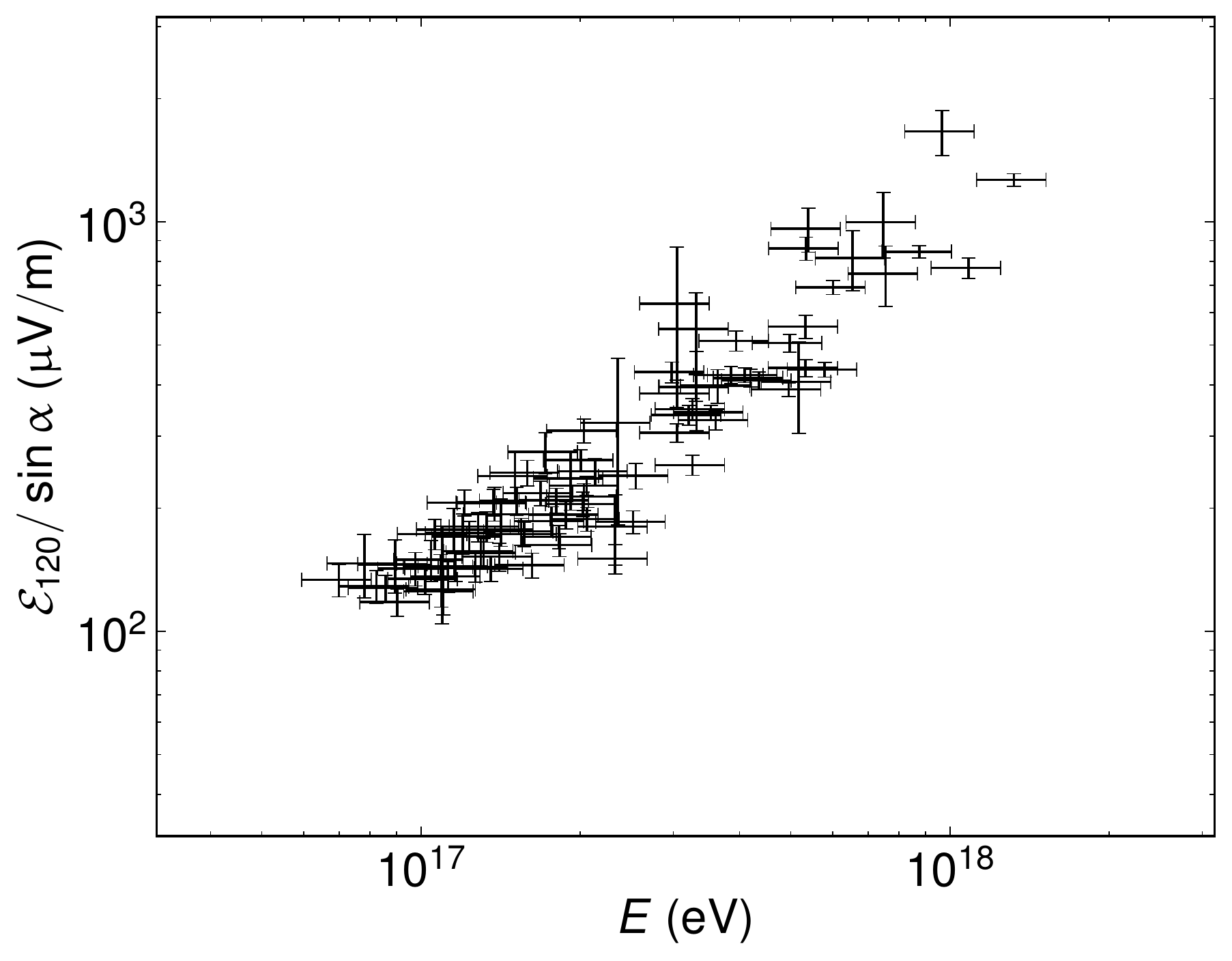}
\caption{The reconstructed electric field strength $\mathcal{E}_{120}$, 
120\,m from the shower axis, normalized for the geomagnetic angle, versus the shower energy reconstructed by Tunka-133.
The error bars here indicate uncertainties from the energy reconstruction of Tunka-133 and the LDF fit.}
\label{fig:ecorrelation}
\end{figure}
For obtaining a simple energy estimator as a first step, one can fit a one-dimensional 
lateral distribution function, like the exponential function,
to the distribution of electric field amplitudes in shower plane coordinates~\cite{LOPESLDF}:
\begin{equation}
\label{eq:LDF}
\mathcal{E} (r) = \mathcal{E}_{r_0}\cdot \exp(-\eta (r-r_0)) .
\end{equation}
We then reconstruct $\mathcal{E}_{120}$, the electric field strength at $r=120$\,m
distance from the shower core~\cite{TRexEreco}. 
To account for the impact of the geomagnetic angle $\alpha$,
$\mathcal{E}_{120}$ is divided by $\sin \alpha$.
In Fig.~\ref{fig:ecorrelation} the resulting quantity is shown as a function of energy reconstructed by Tunka-133.
In this simple form we already obtain a significant correlation and a resolution of 24\%.
However, an accurate reconstruction of the energy as well as of the shower maximum 
has to take finer features of the lateral distribution into account, 
e.g. the azimuthal asymmetry~\cite{TRexEreco}. 
An appropriate analysis is in preparation.

\begin{figure}[tb]
\center
\includegraphics[width=0.7\textwidth]{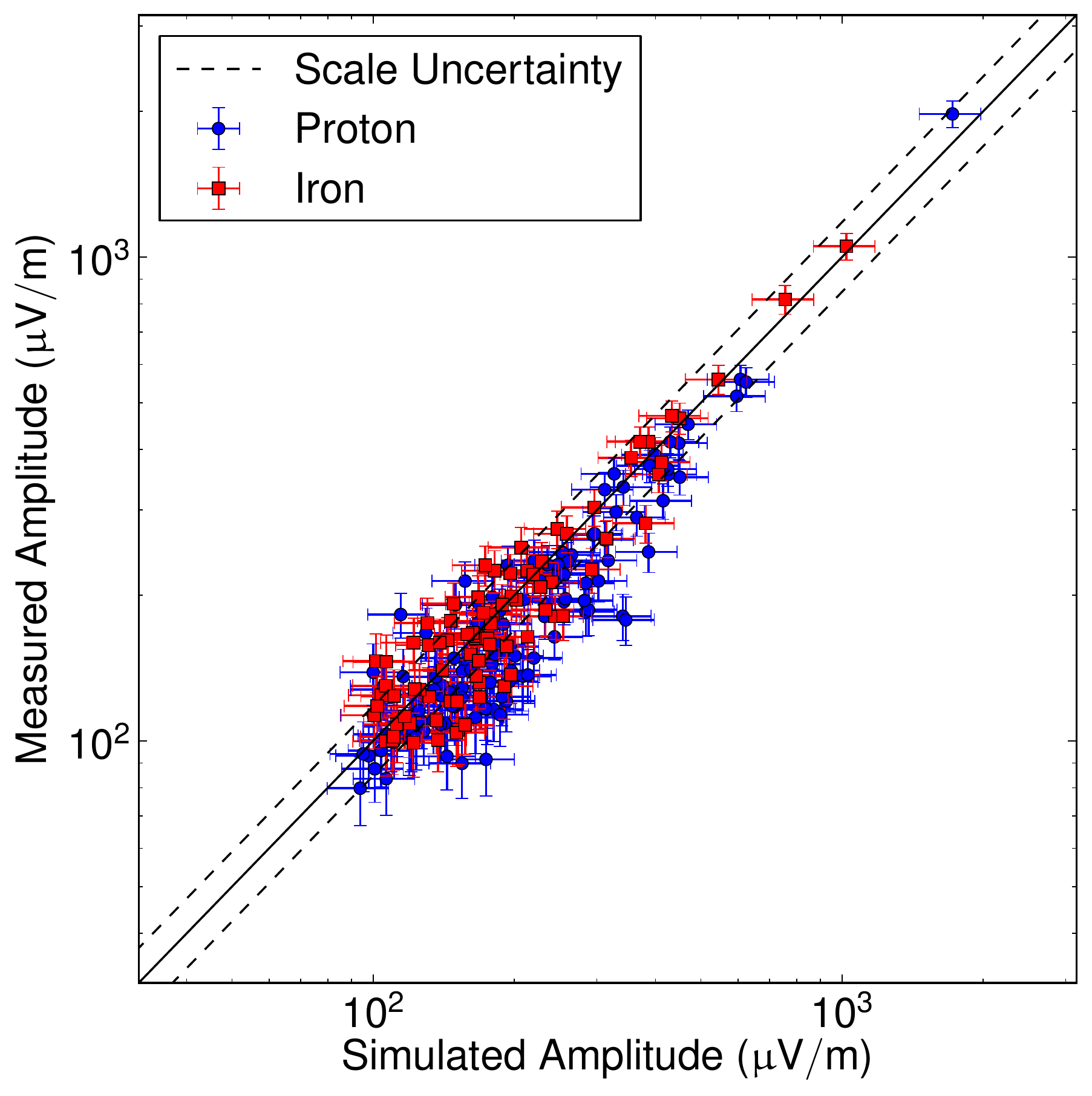}
\caption{Reconstructed amplitudes of radio pulses from stations of Tunka-Rex events 
compared to the amplitudes simulated by CoREAS for iron and proton primaries.
The error bars here indicate uncertainties from the energy reconstruction of Tunka-133, used as input for the simulations, and from the background. Contribution of the scale uncertainty are given in table~\ref{tab:calib} under "Absolute scale".}
\label{fig:coreas}
\end{figure} 
Finally, due to the absolute calibration we can compare our actual measurement 
of the radio emission with model calculations.
Using the Monte-Carlo-Simulation code CoREAS~\cite{CoREAS} with the hadronic interaction model QGSJET-II.04,
we simulated the radio emission of air showers with the same parameters as reconstructed by Tunka-133.
The shower maximum, although it has an impact on the radio signal~\cite{LOPES_xmax},
cannot be set in the simulation, but results from other parameters and
statistical fluctuations.
Therefore, we repeated the simulations several times with different random seeds and 
only take simulated showers into account, 
which happened to have their shower maximum closer than 30\,g/cm$^2$ to the value
reconstructed by Tunka-133. 
These events then undergo a complete detector simulation: 
the response of the antenna and hardware is applied, 
the signal is downsampled and converted to ADC counts according to the 
used data acquisition hardware, and measured noise is added.

Finally, the events are reconstructed the same way as the original events, including the cuts.
This results in 20 iron and 28 proton events which can be used for comparing measurements and simulations.

For each event, we compare the signal of each station. 
Thus, we obtain 83 and 124 amplitude measurements, 
for iron and proton respectively.
In Fig.~\ref{fig:coreas}, the measured signal amplitudes are shown and compared to the simulated amplitudes after adding real measured noise to the simulations.
Over the whole signal range there is a approximately constant factor of 0.96 for iron and 0.84 for proton 
between the measured and simulated signal. 
Consequently, the simulated and measured amplitudes are consistent within the scale uncertainty of the amplitude reconstruction of 17\% (other uncertainties average out). 
Moreover, the small difference between the primaries shows that the radio signal depends only little on the primary particle type.
\section{Conclusion and outlook}
\label{sec:conclusion}
Tunka-Rex started operation in 2012 and successfully measured the radio 
signal of cosmic-ray air showers in coincidence with Tunka-133, an air-Cherenkov detector.
Based on data of the first season, Tunka-Rex was able to confirm the correlation 
of the signal strength with the shower energy and its dependence 
on the geomagnetic angle. Thus, we confirm the picture of a dominant geomagnetic emission mechanism.

An absolute calibration of the signal chain of Tunka-Rex with a calibrated source was performed
with a resulting uncertainty for the absolute amplitude reconstruction of 22\% for individual measurements, including 17\% scale uncertainty.
A comparison of the measured radio emission to air-shower simulations with CoREAS and QGSJET-II.04 revealed
that the measured radio amplitudes can be reproduced by the model within 
the uncertainty.

We recently extended Tunka-Rex by another 19 antennas, connected to the 
scintillator detectors of the new Tunka-Grande array. 
This extension will provide a connection between the scintillator detector and the air-Cherenkov detector,
possibly allowing for an easier cross-calibration of Tunka-133 and Tunka-Grande.
Furthermore, we will also be able to take data during daytime and increase the 
antenna density by a factor of two, enhancing the uptime, as well as the efficiency of Tunka-Rex.

Currently, we are developing methods to reconstruct the primary energy and 
the atmospheric depth of the shower maximum~\cite{TRexEreco}.
Testing these on the still blinded data set of the 2013/2014 measurement season, 
we will determine the achievable precision 
of the radio detection technique.
Provided a sufficient precision, Tunka-Rex can dramatically enhance the total 
statistics of the Tunka-133 measurement around 10$^{18}$\,eV exploiting the new scintillator trigger for day-time measurement.



\section*{Acknowledgment}
We would like to thank many persons for supporting Tunka-Rex and for helping 
in the production and deployment, in particular N.~Barenthien, F.~Bocci, H.~Bolz, H.~Bozdog, L.~Classen, 
T.~Fischer-Wasels, V.~Lenok, J.~Oertlin, P.~Reiher, P.~Satunin, V.~Savinov,
C.~Spiering and A.~Tokareva.
Tunka-Rex has been funded by the German Helmholtz association (grant HRJRG-303) and 
supported by the Helmholtz Alliance for Astroparticle Physics (HAP), as well as KCETA.
This work was supported by the Russian Federation Ministry of Education and Science
(agreement 14.B25.31.0010, zadanie 3.889.2014/K) and
the Russian Foundation for Basic Research 
(Grants 12-02-91323, 13-02-00214, 13-02-12095, 14002-10002).


\bibliographystyle{elsarticle-num}
\bibliography{NIMA_ref}


\end{document}